\documentclass[twocolumn,preprintnumbers,amsmath,amssymb,groupedaddress,superscriptaddress,prl]{revtex4-2}

\usepackage[utf8]{inputenc}
\usepackage{graphicx}
\usepackage{dcolumn}
\usepackage[colorlinks=true,citecolor=blue,urlcolor=blue]{hyperref}
\usepackage[dvipsnames]{xcolor}
\usepackage{subfigure}
\usepackage{ulem} 
\begin{document}
\title{Dynamical formation of multiple quantum droplets in a Bose-Bose mixture}

\author{L. Cavicchioli}\email{cavicchioli@lens.unifi.it}
\affiliation{Istituto Nazionale di Ottica, CNR-INO, 50019 Sesto Fiorentino, Italy}
\affiliation{\mbox{Dipartimento di Fisica e Astronomia and LENS, Universit\`{a}
di Firenze, 50019 Sesto Fiorentino, Italy}}

\author{C. Fort}\email{chiara.fort@unifi.it}
\affiliation{\mbox{Dipartimento di Fisica e Astronomia and LENS, Universit\`{a}
di Firenze, 50019 Sesto Fiorentino, Italy}}
\affiliation{Istituto Nazionale di Ottica, CNR-INO, 50019 Sesto Fiorentino, Italy}

\author{F. Ancilotto}
\affiliation{\mbox{Dipartimento di Fisica e Astronomia 'Galileo Galilei' and
CNISM, Universit\`{a} di Padova, 35131 Padova, Italy}}
\affiliation{\mbox{CNR-Officina dei Materiali (IOM), via Bonomea, 265 - 34136 Trieste, Italy}}

\author{M. Modugno}
\affiliation{Department of Physics, University of the Basque Country UPV/EHU,
  48080 Bilbao, Spain}
\affiliation{IKERBASQUE, Basque Foundation for Science, 48013 Bilbao, Spain}
\affiliation{EHU Quantum Center, University of the Basque Country UPV/EHU,
  Leioa, Biscay, Spain}

\author{F. Minardi}
\affiliation{Dipartimento di Fisica e Astronomia, Universit\`{a} di Bologna,
40127 Bologna, Italy}
\affiliation{Istituto Nazionale di Ottica, CNR-INO, 50019 Sesto Fiorentino, Italy}
\affiliation{\mbox{Dipartimento di Fisica e Astronomia and LENS, Universit\`{a}
di Firenze, 50019 Sesto Fiorentino, Italy}}

\author{A. Burchianti}
\affiliation{Istituto Nazionale di Ottica, CNR-INO, 50019 Sesto Fiorentino,
  Italy}
\affiliation{\mbox{Dipartimento di Fisica e Astronomia and LENS, Universit\`{a}
di Firenze, 50019 Sesto Fiorentino, Italy}}

\begin{abstract}
  We report on the formation of multiple quantum droplets in a heteronuclear $^{41}$K-$^{87}$Rb mixture released in an optical waveguide. By a sudden change of the interspecies interaction from the non-interacting to the strongly attractive regime, we initially form a single droplet in an excited compression-elongation mode. The latter axially expands up to a critical length and then splits into two or more smaller fragments, recognizable as quantum droplets. We find that the number of formed droplets increases with decreasing interspecies attraction and increasing atom number. We show, by combining theory and experiment, that this behavior is consistent with capillary instability, which causes the breakup of the stretching droplet due to the surface tension. Our results open new possibilities to explore the properties of quantum liquids and systems of multiple quantum droplets in two-component bosonic mixtures.
\end{abstract}

\maketitle

Quantum droplets are a novel liquid state at low density that forms in degenerate ultracold gases with competing interactions \cite{Barbut_2019, Malomed_2020, Böttcher_2021, Reimann_2023}. In bosonic mixtures, such states emerge as a result of the balance between the attractive mean-field energy and the repulsive Lee-Huang-Yang (LHY) correction due to quantum fluctuations \cite{Petrov2015, PetrovAstrakharchik}. So far quantum droplets have been observed in spin mixtures of $^{39}$K \cite{Cabrera2018, Semeghini2018},  heteronuclear mixtures of $^{41}$K-$^{87}$Rb \cite{Derrico2019,Burchianti2020} and $^{23}$Na-$^{87}$Rb  \cite{Guo2021}, and in single-species dipolar gases \cite{kadau2016,FerrierBarbut2016,Schmitt2016,Ferlaino2016,FerrierBarbutJPB2016,Wenzel2017}. In the latter case, where contact repulsion is counteracted by dipole–dipole attraction, the droplets are elongated along the dipole direction and may arrange in regular arrays under confinement \cite{FerrierBarbut2016,Schmitt2016,Wenzel2017}.
Conversely, in atomic mixtures, where only contact interactions are at play, the system is expected to form a single self-bound droplet, with a fixed density ratio of the two components. These binary droplets are a prime example of an isotropic quantum liquid and have peculiar properties like the predicted self-evaporation mechanism \cite{Petrov2015, ferioli2019dynamical, FortModugno2021}. In experiments, their liquid-like behavior has been probed by studying the free-space collisions between $^{39}$K droplets \cite{Ferioli2019_collision}.
The formation of binary droplets has been also reported in quasi-1D \cite{Cheiney2018,Derrico2019} and quasi-2D traps \cite{Cabrera2018}; in the former case a smooth crossover between bright solitons and droplets has been observed \cite{Cheiney2018}.

Despite binary droplets having attracted much interest, the study of their dynamics is hindered by the effect of three-body collisions, which shortens the droplet lifetime. However, quantum droplets composed by $^{41}$K and $^{87}$Rb can still survive for several tens of milliseconds \cite{Derrico2019,Burchianti2020}, opening new possibilities to study the elusive properties of these quantum liquids.

\begin{figure*}[t!]
  \centering
  \includegraphics[width=.90\linewidth]{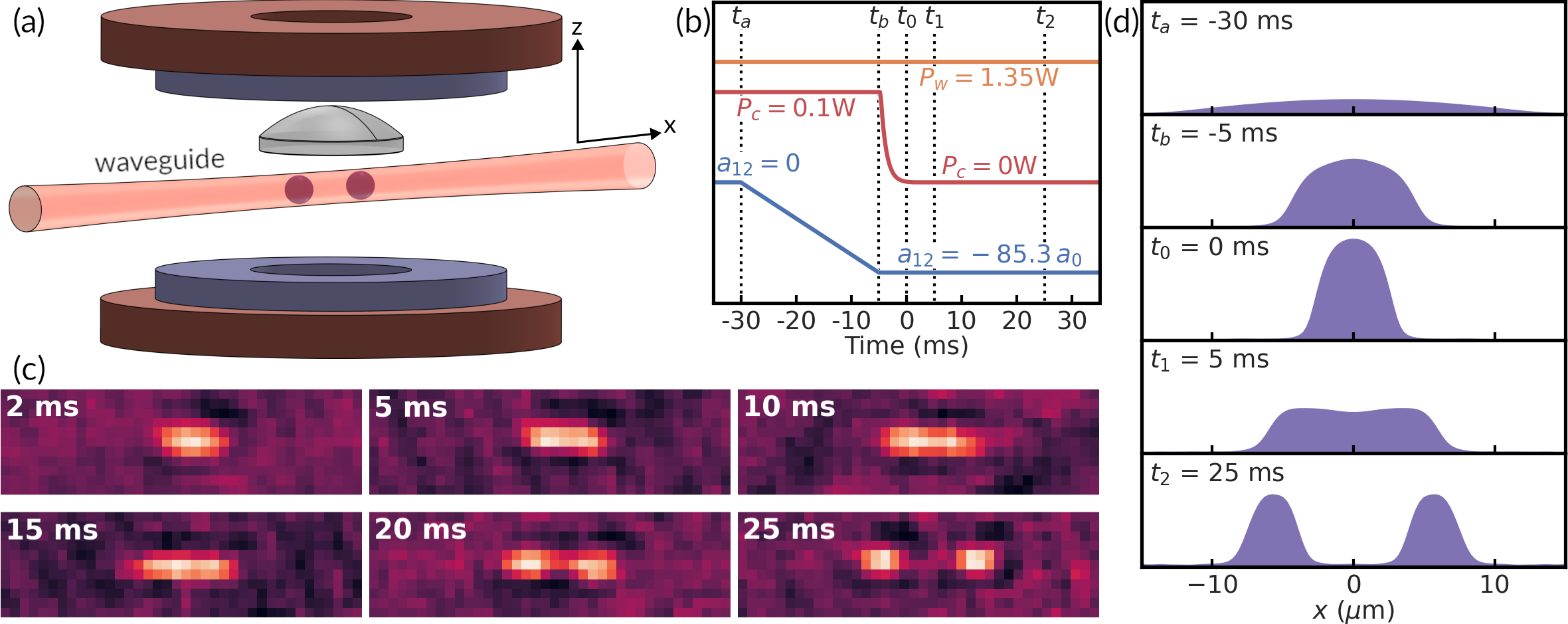}
  \caption{\label{fig:fig1}(a) Schematics of the experiment, showing the “waveguide” beam, the Feshbach and the quadrupole coils (in brown and blue, respectively) and the objective lens. (b) Experimental sequence used to trigger the out-of-equilibrium dynamics where P$_w$ and P$_c$ indicates the power of the “waveguide” and the “crossed” beam, respectively. (c) Examples of \textit{in situ} absorption images of $^{41}$K in the $xy$ plane ($57\;\mu$m $\times$ $15\;\mu$m)  after release in the waveguide with $a_{12}=-85.3(1.5) a_0$. Images taken at times 2, 5, 10, 15, 20, 25 ms. Each image is normalized to its maximum value to increase the visibility.
    (d) Axial density profiles of $^{41}$K from GP simulations, for $N_1=N_2=4 \times 10^4$ and $a_{12}=-85.3 a_0$, at the times $t_a$, $t_b$, $t_0$, $t_1$ and $t_2$ (from top to bottom) indicated in the experimental sequence (b).}
\end{figure*}

In this Letter, we show that a single $^{41}$K-$^{87}$Rb droplet, released in an optical waveguide, undergoes a dynamical instability, which results in the formation of multiple droplets. Due to the confinement, the droplet ground state would take the form of a cylindrical filament, whose length increases for increasing atom numbers or decreasing interspecies attraction. We trigger the instability by preparing the system out of equilibrium, through a sudden change of the interspecies interaction from the non-interacting to the strongly attractive regime. This excites a compression-elongation motion, mainly in the direction of the waveguide. We track the evolution of the droplet density distribution by \textit{in situ} absorption imaging. The density profiles show that, once the length of the axially stretched droplet exceeds a critical value, it breaks up into two, or more, smaller droplets. The observed dynamics is well reproduced by numerical simulations, based on two coupled Gross–Pitaevskii (GP) equations at $T=0$, including the LHY correction for heteronuclear mixtures \cite{Ancilotto2018,Minardi2019}. We study both experimentally and theoretically the outcome of the droplet breakup process by varying the interspecies attraction and the atom number. We explain our results in terms of the Plateau–Rayleigh (PR) capillary instability of classical inviscid \cite{Plateau1857,Rayleigh1879} and quantum liquid filaments \cite{Speirs2020,Ancilotto2023}. According to this, the initial droplet breaks up into sub-droplets due to surface tension, which acts to minimize the surface area.

We start the experiment with a non-interacting dual-species condensate of $^{41}$K and $^{87}$Rb, both prepared in their hyperfine ground state \cite{Burchianti2020, Cavicchioli2022}, with a total atom number $N$ of about $ 4 \times 10^5$ and a population imbalance $P={(N_1-N_2)}/{N}\sim -0.4$  (hereafter we use the notation: $1 \rightarrow$ $^{41}$K and $2 \rightarrow$ $^{87}$Rb). The mixture is initially confined in a cigar-shape optical trap formed by two laser beams at 1064~nm: a primary “waveguide” beam [Fig.~\ref{fig:fig1}(a)],  along the $x$ axis, and an auxiliary “crossed” beam forming an angle of 45$^\circ$ with the waveguide in the $xy$ plane \cite{supp}. The interspecies scattering length $a_{12}$ is adjusted by a vertical homogeneous magnetic field $B_z$, tuned around 72~G, corresponding to the zero-crossing in between two Feshbach resonances \cite{Thalhammer2008}. In the range where $B_z$ is varied, the intraspecies scattering lengths are almost constant: $a_{11}= 62.0 a_0$ \cite{Bighin_PRA2022} and $a_{22}=100.4 a_0$ \cite{Marte2002}. The differential gravitational sag, due to the different trap frequencies of $^{41}$K and $^{87}$Rb, is compensated with a vertical magnetic gradient $b_z=-16.6$~G/cm \cite{Burchianti2020, Cavicchioli2022}, produced by a quadrupole magnetic field. Here, we excite the dynamics by linearly decreasing $a_{12}$ from zero to $-85.3(1.5) a_0$ \footnote{For $a_{12} < -73.6a_0$, the mixture is stabilized by the LHY term and turns into a quantum droplet above a critical atom number \cite{Petrov2015}.}, in $t_b-t_a=$25~ms, corresponding to a variation of $B_z$ with a constant rate of $-0.22$~G/ms. Then, we exponentially ramp down the crossed beam in 5~ms [sequence sketched in Fig.~\ref{fig:fig1}(b)]. The final potential experienced by $^{41}$K ($^{87}$Rb) is approximately harmonic with a radial average frequency of $136(3)$~Hz [$100(2)$~Hz] and an axial frequency of $3.4(2)$~Hz [$1.9(1)$~Hz], the latter being mainly due to the magnetic curvature of the Feshbach field.
Starting from $t_0=0$, we follow the system dynamics through the sequential imaging of $^{41}$K and $^{87}$Rb, in the $xy$ plane, using an objective with a measured resolution of $1.5(1)\mu$m ($1/e^2$ Gaussian width) \cite{supp}. The droplet shape is determined directly from the \textit{in situ} absorption images of $^{41}$K, the minority component, which is entirely bound \footnote{The ratio  of $^{41}$K and $^{87}$Rb atom numbers in the droplet is approximately constant and close to one \cite{Derrico2019}. We have typically $N_1 \sim 0.4  \ N_2$, this implies that all $^{41}$K is bound while a residual fraction of $^{87}$Rb remains unbound.}. After the $^{41}$K imaging pulse, the bound state is dissociated and starts to expand. Thus, the subsequent $^{87}$Rb imaging cannot provide information about the droplet size and is used as monitor signal. We observe the following [Fig.~\ref{fig:fig1}(c)]: the droplet axially expands \footnote{The system's center of mass moves along the $x$ axis, mainly due to a shift between the initial droplet position and the center of the magnetic curvature. The images in Fig.~\ref{fig:fig1}c are  re-centered on the center of mass.} up to approximately 15~ms  and then splits into two smaller fragments which move apart. To clarify the nature of these clusters, we have followed their fall in free space. Once the waveguide is abruptly switched off, their size remains constant up to 10~ms, allowing us to recognize them as self-bound quantum droplets \cite{supp}.
In Fig.~\ref{fig:fig1}(d) we also report the simulated axial density profiles of $^{41}$K, obtained by solving two coupled time-dependent GP equations 
\cite{supp}, corresponding to different times of the experimental sequence, starting from the non-interacting regime. As $a_{12}$ is ramped down to the attractive side, the sample initially shrinks in size and then, after $t_0$, stretches along the waveguide axis until it breaks up, reproducing the observed dynamics.

\begin{figure}[t!]
  \centering
  \includegraphics[width=0.75\columnwidth]{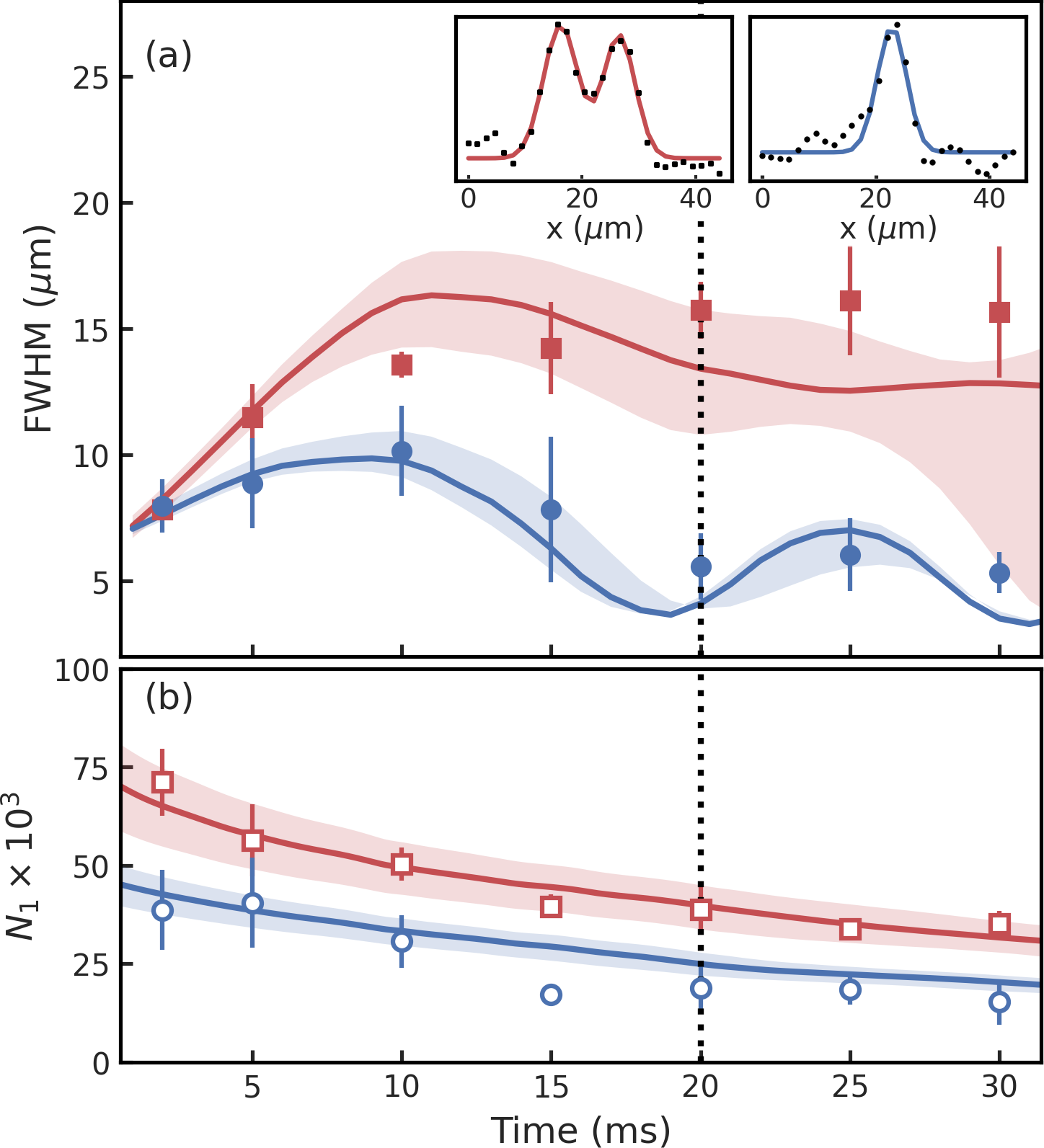}
  \caption{\label{fig:fig2}(a) Evolution of the axial droplet size (FWHM) and (b) the $^{41}$K atom number $N_1$, in the waveguide, for $a_{12}=-85.3(1.5) a_0$. The red squares and the blue circles correspond to a ramp of $a_{12}$ lasting 25~ms and 100~ms, respectively. In the first case the droplet splits in two (the vertical dotted line indicates the breakup time, see text). In the second case, the droplet undergoes a compression-elongation mode without breaking up. In the inset we show two typical density profiles recorded at 20~ms: (i) breaking and (ii) no breaking. The error bars correspond to a standard deviation of typically five independent measurements. The lines represent the corresponding quantities extracted from  GP simulations,  with atom numbers $N_1=1.2\times 10^5$ and $N_2=3.0\times 10^5$ at the time $t_a$ before the interaction ramp to $a_{12}=-85.3a_0$ and a three-body loss coefficient $K_3=7\times 10^{-41}$m$^6/$s \cite{Derrico2019}. The shaded areas indicate the systematic relative uncertainty of the experimental atom number of 20\%.}
\end{figure}

In Fig.~\ref{fig:fig2} we show the droplet axial size (full width at half maximum, FWHM) extracted from a Gaussian fit of the experimental density profiles, and the measured $^{41}$K atom number, $N_1$, as a function of time for the case reported in Fig.~\ref{fig:fig1} (filled and empty red squares). Starting from the breakup time (the vertical dotted line) the density profile is fitted with a double Gaussian function  [inset (i) in Fig.~\ref{fig:fig2}(a)] and, thereafter we plot $d+\sum_{j} \delta_j$, with $d$ the distance between the center of the two Gaussian, and $\delta_j$ the half width at the half maximum (HWHM) of the $j$-th droplet. The atom number decreases during the dynamics due to the effect of three-body losses [see Fig.~\ref{fig:fig2}(b)] with a decay time that is consistent with our previous estimation of the $K_3$ coefficient \cite{Derrico2019}. In Fig.~\ref{fig:fig2} we also report the corresponding observables (filled and empty blue circles) obtained when the dynamics is triggered by linearly ramping $a_{12}$ such that $t_b-t_a=$100~ms, without changing the ramp endpoints. This has the two-fold effect of enhancing the atomic losses and decreasing the oscillation amplitude of the droplet size \footnote{The critical ramping time for the onset of instability depends on $a_{12}$ and $N_1$. In the experiment, we have observed that a 25~ms interaction ramp allows to explore the droplets breakup dynamics over a wide range of interactions.}.
Consequently, the droplet does not reach the critical length for the onset of the instability, and a single density peak [inset (ii) in Fig.~\ref{fig:fig2}(a)] is observed during the evolution. The solid lines in Fig.~\ref{fig:fig2}, which qualitatively agree with the experimental data, are the results of GP simulations performed without free parameters including three-body losses, the initial population imbalance, and the effective potentials \cite{supp}.  Once taking into account the experimental uncertainty on the atom number (shaded areas), we  find that, in the case of breaking, the partially separated droplets may merge again in the simulation at later times for low atom numbers.

\begin{figure}[t!]
  \centering
  \includegraphics[width=\columnwidth]{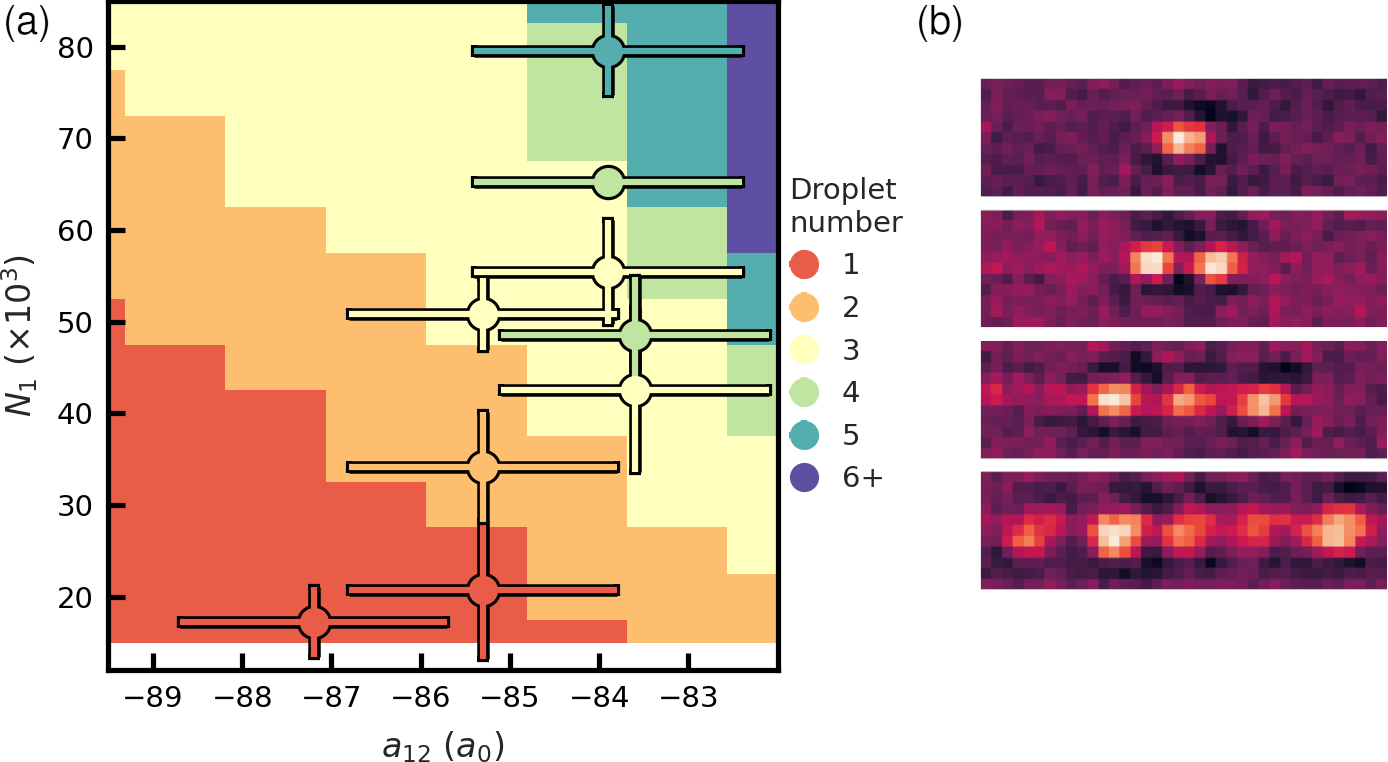}
  \caption{\label{fig:fig3}(a) Comparison between experimental results and GP simulations. The circle points represents the number of droplets  measured in the experiment as a function of $a_{12}$ and $N_1$. Vertical error bars are a standard deviation of typically five independent measurements while horizontal bars show the uncertainty in the value of $a_{12}$. The colored map represents the regions in the ($a_{12},N_1$) plane where the GP simulations show $1,2,3,\dots$ droplets. Experimental data and theory follow the same color legend. (b) Typical \textit{in situ} absorption images of $^{41}$K in the $xy$ plane ($57\;\mu$m $\times$ $15\;\mu$m) taken at a time of 30 ms for decreasing $|a_{12}|$ and increasing $N_1$ (from top to bottom).  Each image is normalized to its maximum value to enhance the visibility.}
\end{figure}

We repeated the experiment by varying the endpoint value of $a_{12}$ in the interaction ramp lasting 25~ms. Due to the three-body losses, a change of $a_{12}$, and thus of the droplet density, also affects the atom number: $N_1$ increases by decreasing $\left | a_{12} \right  |$. Within the parameter space accessible by the experiment, we find that, once the instability sets in, the droplet splits either in two or more fragments, whose number grows for weaker interspecies attractions and larger atom numbers. This behavior is illustrated in Fig.~\ref{fig:fig3}(a), where we plot the number of droplets (circle points) counted in the absorption images at a time ranging from 20 to 30 ms,  as a function of $a_{12}$ and $N_1$. A set of images, corresponding to different outcomes of the droplet dynamics, is shown in Fig.~\ref{fig:fig3}(b). The experimental results are well reproduced by GP simulations [colored map in Fig.~\ref{fig:fig3}(a)] performed assuming $N_1=N_2$, neglecting three-body losses, and a simplified form of the waveguide potential. We have verified, in selected cases, that these approximations do not significantly impact the results \cite{supp}.

Our findings are reminiscent of capillary instability, which occurs in a variety of physical systems, including ordinary liquids, superfluid and normal $^4$He and nuclear matter \cite{Eggers_2008}. This phenomenon has also been predicted in bosonic mixtures, both in the droplet \cite{Ancilotto2023} and in the immiscible regime \cite{Sasaki2011}. In the former case, it has been demonstrated that a $^{41}$K-$^{87}$Rb filament is unstable to any varicose perturbation \footnote{For a filament along the $x$ axis, the classical PR instability assumes a perturbation of the form $R(x,t)=R_0+\delta e^{\omega (t-t_0)} \cos(kx)$, with $k=2 {\pi/ \lambda}$  the wave vector, $\delta$ the amplitude of the perturbation, and $\omega(k)$ the growth rate.} with wavelength $\lambda$ approximately larger than $2 \pi R_0$, with $R_0$ the filament radius. In close analogy with classical PR instability, the most unstable mode, which eventually breaks up the thread, corresponds to $\lambda_c \simeq 9R_0$. The relevant time scale of the instability is given by the capillary time:
\begin{equation}
  \label{eq:tauc}
  \tau_c= \sqrt{\frac{(n_1 m_1+n_2 m_2) R_0^3}{\mathcal{T}_0}},
\end{equation}
where $n_i$ and $m_i$ are the density and the mass of the $i$-th atomic species, respectively, and $\mathcal{T}_0$ is the surface tension of the mixture \cite{Boronat2021}. Within this model, once the length of the stretching droplet approaches $\lambda_c$ (at $t=t^*$), the corresponding mode, with an initial amplitude $\delta$, starts to grow exponentially with a rate given by $(2.9\tau_c)^{-1}$, according to the PR spectrum \cite{Eggers_2008}. Following \cite{Driessen2013}, we define the breaking time $t_\text{break}$ as the instant when the mode amplitude reaches the radial size of the filament $R_0$, causing it to break up into droplets. We thus obtain $t_\text{break}=t^*+(2.9\tau_c)\ln(R_0/\delta)$. The number $N_d$ of daughter droplets is then determined by the length of the mother droplet $L_\mathrm{max}$ at $t_\mathrm{break}$, according to the relation $N_d=1+\lfloor L_\mathrm{max}/\lambda_c-1/2 \rfloor$.

In order to link the observed behavior to PR instability, we extracted both $L_\mathrm{max}$ and $t_\mathrm{break}$ from the systematic GP simulations shown in Fig. 3 (see \cite{supp} for the criteria used to define  these two parameters).
These quantities are
leveraged for a comparison with the predictions of the capillary instability model of an inviscid liquid filament.

\begin{figure}[t!]
  \centering
  \includegraphics[width=0.95\columnwidth]{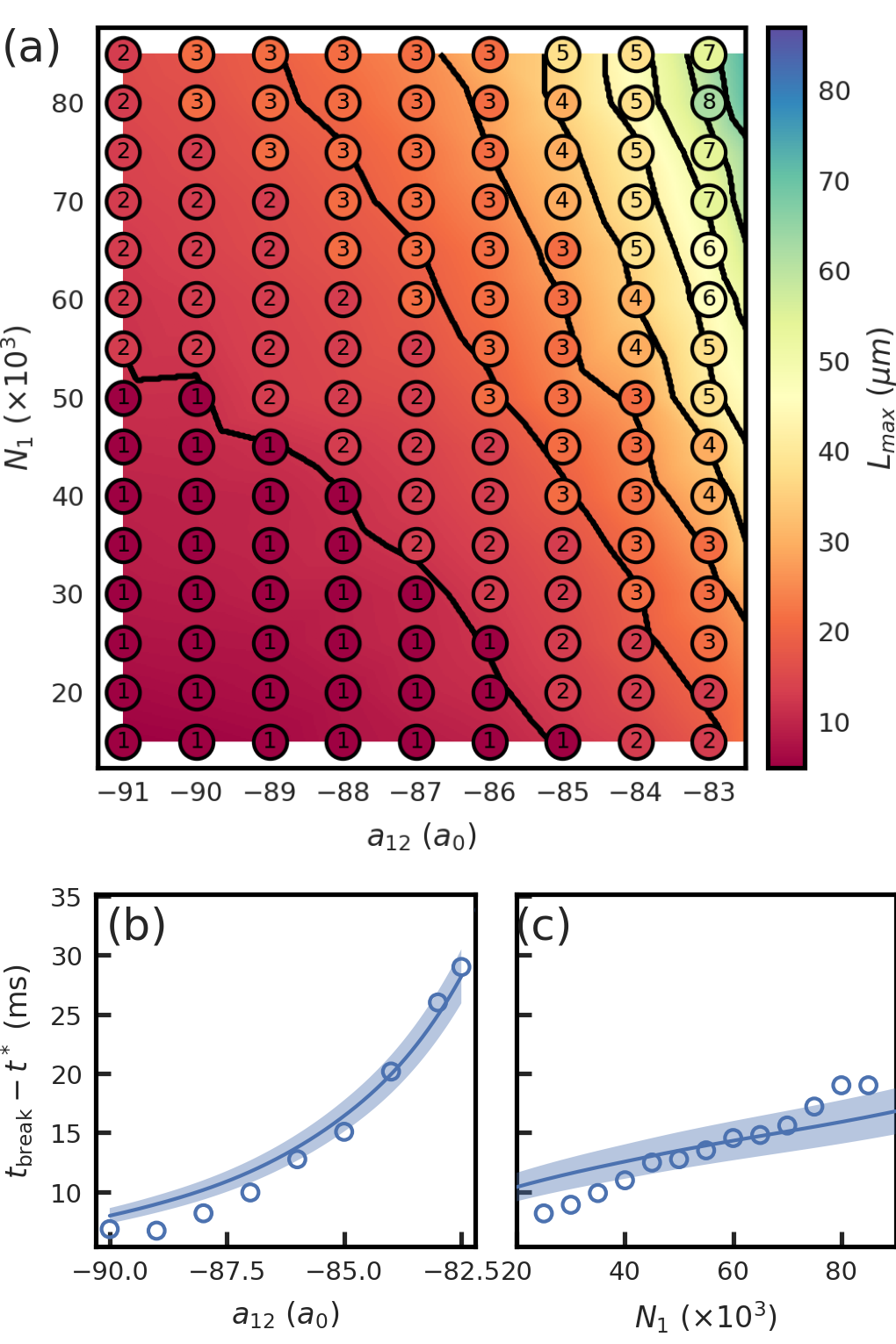}
  \caption{\label{fig:fig4} Results of GP simulations. (a) Map of $L_\mathrm{max}$ (color plot) and $N_d$ (labeled points)  for different ($a_{12},\,N_1$). The contour lines represent $L_\mathrm{max}=(n+1/2)\,\lambda_c$ with $n \,\in \mathbb{Z}^+$. (b) Values of $t_\mathrm{break}-t^*$  (data points) as a function of $a_{12}$ for $N_1=5\times 10^4$ and (c) as a function of $N_1$ for $a_{12}=-86a_0$. The solid lines are the corresponding fitting curves using Eq.~\eqref{eq:tauc} (shaded areas represent prediction bands with 99\% confidence level).}
\end{figure}

Fig.~\ref{fig:fig4}a shows $L_\mathrm{max}$ as a function of $a_{12}$ and $N_1$. As the interspecies attraction weakens and the atom number increases, the filament undergoes greater elongation before breaking. The contour lines in Fig.~\ref{fig:fig4}a represent $L_\mathrm{max}=1.5 \lambda_c, 2.5 \lambda_c, 3.5 \lambda_c,\dots$, with $\lambda_c=9 R_0$ and $R_0$ given by the droplet radial width $\sigma_r\simeq 0.85$~$\mu$m, as determined by the waveguide potential \cite{supp}. These lines bound the regions where, due to the PR instability, one expects $N_d=1,2,3,\dots$.
The labeled points show the number of droplets formed in the simulations after filament breakage. The corresponding experimental data points, already shown in Fig.~\ref{fig:fig3}, are not reported here to facilitate the map readability. The effective number of daughter droplets aligns well with the PR instability predictions. We emphasize that, while $L_{max}$ is determined by the chosen experimental sequence used to initiate the dynamics, $N_d$ depends solely on the ratio between $L_{max}$ and $\lambda_c$, as expected from capillary instability  \cite{supp}.

We also compare the breaking time with the capillary time estimated by Eq.~\eqref{eq:tauc}, assuming as $n_i$ the density of the filament at $t^*$ \footnote{
  In the range of parameters used in the simulations, if the length of the filament is smaller than $\lambda_c$ before expanding, then we find $t^* \in [-2,0.5]$ ms, else we set $t^*=0$ ms.}
and as $\mathcal{T}_0$ the surface tension calculated according to \cite{Boronat2021}.  In Fig.~\ref{fig:fig4}b (Fig.~\ref{fig:fig4}c) the data points are the simulated values of $t_\mathrm{break}-t^*$ with $N_1=5\times 10^4$ ($a_{12}=-86a_0$) as a function of $a_{12}$ ($N_1$). The solid lines are obtained from a fit using $2.9 \ln(R_0/\delta) \tau_c(a_{12},N_1)$ as the fitting function, with $\delta$ as the only fitting parameter. The fit returns $\delta/R_0=0.116\pm0.06$ and $\delta/R_0=0.121\pm0.06$ for the cases in Fig.~\ref{fig:fig4}b and \ref{fig:fig4}c, respectively. These values agree within the error bar and, since $\delta \ll R_0$ , justify the used  linear
approximation for the instability  \cite{Driessen2013}. We can conclude that the observed scaling of the breakup time with $a_{12}$ and $N_1$ is qualitatively consistent with the capillary time scaling, assuming a constant
excitation amplitude $\delta/R_0$.
We point out that in Fig.~\ref{fig:fig4}b, $\tau_c$ depends on $a_{12}$ only through $\mathcal{T}_0$ since the atomic density is fixed by the filament length $\lambda_c$, whereas in Fig.~\ref{fig:fig4}c $\tau_c$ depends on $N_1$ only through the atomic density since $\mathcal{T}_0$ here is fixed by $a_{12}$.

Summarizing, the number of droplets resulting from the breakup of the atomic filament is proportional to $L_\mathrm{max}$ [Fig.~\ref{fig:fig4}(a)] and the breakup time to $\tau_c$ [Figs.~\ref{fig:fig4}(b) and \ref{fig:fig4}(c)], in agreement with the PR model.

In conclusion, we have studied the fate of a binary quantum droplet created out of equilibrium and released in a waveguide. The  droplet evolves in a filament that, above a critical length, breaks up into multiple quantum droplets due to capillary instability.
Our results provide deep insight into the liquid-like behavior of binary quantum droplets and direct evidence of their surface tension. This work enriches the variety of hydrodynamic instabilities reported in ultracold gases, in close analogy with what observed in classical fluids \cite{Hernández-Rajkov_2024, huh2024,Spielman_2024}. Future directions may include the study of surface modes, the role of confinement in the splitting and merging dynamics, and  the emergence of quantum effects in such processes, as already observed in experiments with helium droplets \cite{Vicente2000,Ishiguro_2004}.
Finally, the ability  to realize multi-droplet arrays paves the way for future studies aimed to address the coherence properties of such states. Indeed, recent theoretical works \cite{Reimann2022,Ancilotto2023} suggest that, in imbalanced mixtures, a superfluid background may impart supersolid properties to the droplets.
\\
\begin{acknowledgments}
  We thank Marco Fattori for critical reading, and the LENS Quantum Gases group for discussions. This work was supported by the Italian Ministry of University and Research under the PRIN2022 project
  No. 20227JNCWW and, co-funded the European Union–Next Generation EU under the PNRR MUR project PE0000023-NQSTI and under the ‘Integrated infrastructure initiative in Photonic and Quantum Sciences’ I-PHOQS. MM acknowledges support from Grant No. PID2021- 126273NB-I00 funded by MCIN/AEI/10.13039/501100011033 and by “ERDF A way of making Europe,” and from the Basque Government through Grant No. IT1470-22.
\end{acknowledgments}
\bibliographystyle{apsrev4-2}
\bibliography{biblio}
\end{document}


\title{Supplemental Material to: Dynamical formation of multiple quantum
droplets in a Bose-Bose mixture}

\author{L. Cavicchioli}\email{cavicchioli@lens.unifi.it}
\affiliation{Istituto Nazionale di Ottica, CNR-INO, 50019 Sesto Fiorentino, Italy}
\affiliation{\mbox{Dipartimento di Fisica e Astronomia and LENS, Universit\`{a}
di Firenze, 50019 Sesto Fiorentino, Italy}}

\author{C. Fort}\email{chiara.fort@unifi.it}
\affiliation{\mbox{Dipartimento di Fisica e Astronomia and LENS, Universit\`{a}
di Firenze, 50019 Sesto Fiorentino, Italy}}
\affiliation{Istituto Nazionale di Ottica, CNR-INO, 50019 Sesto Fiorentino, Italy}

\author{F. Ancilotto}
\affiliation{\mbox{Dipartimento di Fisica e Astronomia 'Galileo Galilei' and
CNISM, Universit\`{a} di Padova, 35131 Padova, Italy}}
\affiliation{\mbox{CNR-Officina dei Materiali (IOM), via Bonomea, 265 - 34136 Trieste, Italy}}

\author{M. Modugno}
\affiliation{Department of Physics, University of the Basque Country UPV/EHU,
48080 Bilbao, Spain}
\affiliation{IKERBASQUE, Basque Foundation for Science, 48013 Bilbao, Spain}
\affiliation{EHU Quantum Center, University of the Basque Country UPV/EHU,
Leioa, Biscay, Spain}

\author{F. Minardi}
\affiliation{Dipartimento di Fisica e Astronomia, Universit\`{a} di Bologna,
40127 Bologna, Italy}
\affiliation{Istituto Nazionale di Ottica, CNR-INO, 50019 Sesto Fiorentino, Italy}
\affiliation{\mbox{Dipartimento di Fisica e Astronomia and LENS, Universit\`{a}
di Firenze, 50019 Sesto Fiorentino, Italy}}

\author{A. Burchianti}
\affiliation{Istituto Nazionale di Ottica, CNR-INO, 50019 Sesto Fiorentino,
Italy}
\affiliation{\mbox{Dipartimento di Fisica e Astronomia and LENS, Universit\`{a}
di Firenze, 50019 Sesto Fiorentino, Italy}}

\begin{abstract}
\end{abstract}

\maketitle

\section{Experimental Methods}
\subsection{\label{sec:level1} Trapping potentials}

The optical dipole trap, where we produce the initial double condensate, is
formed by two crossed  beams derived from the same laser source at 1064 nm.
Their power is independently controlled by means of separate acousto-optic
modulators. The two beams intersect at an angle of $45^\circ$ in the horizontal
$xy$  plane. The ``waveguide''  beam is directed along the $x$ axis (Fig.
\ref{fig:FigSM1}), orthogonal to the gravity direction $z$ (the angle with the
horizontal plane is smaller than $0.05^\circ$ and is neglected throughout), and
it has a waist $w_w=95(5)$~$\mu$m and a  power $P_w=1.35$~W, while the auxiliary
``crossed beam'' has a waist $w_c=76(4)$~$\mu$m and a power $P_c=0.1$~W.
The  trap, including the magnetic potentials and gravity, is approximately
harmonic with  estimated frequencies of about $(44, 166, 145)$ and $(32, 122, 106)$~Hz for
$^{41}$K  and $^{87}$Rb, respectively. 
At the time $t=0$ (see Fig.~1(b) in the
main text), when the expansion leading to the droplet breaking starts, $P_c$ is
equal to zero and the remaining harmonic confinement frequencies are
approximately $(3.4, 153, 116)$ and $(1.9, 112, 85)$ Hz,  for $^{41}$K and
$^{87}$Rb, respectively. The axial harmonic confinement is mainly due to the
curvature of the Feshbach field, $C \simeq 5$ G/cm$^2$. 

After the release in the waveguide, in addition to the mentioned elongation, the
droplet displays also an axial motion of the center of mass, which is mainly due
to the shift between the Feshbach-field symmetry axis and the starting position
of the droplet ($\sim 250 \mu$m). This produces an overall displacement of the
droplets center of mass of about 60 $\mu$m  after an evolution time of 30 ms. 

\subsection{\label{sec:level2} High-resolution imaging}

\begin{figure}[b!]
  \begin{center}
    \includegraphics[width=1\columnwidth]{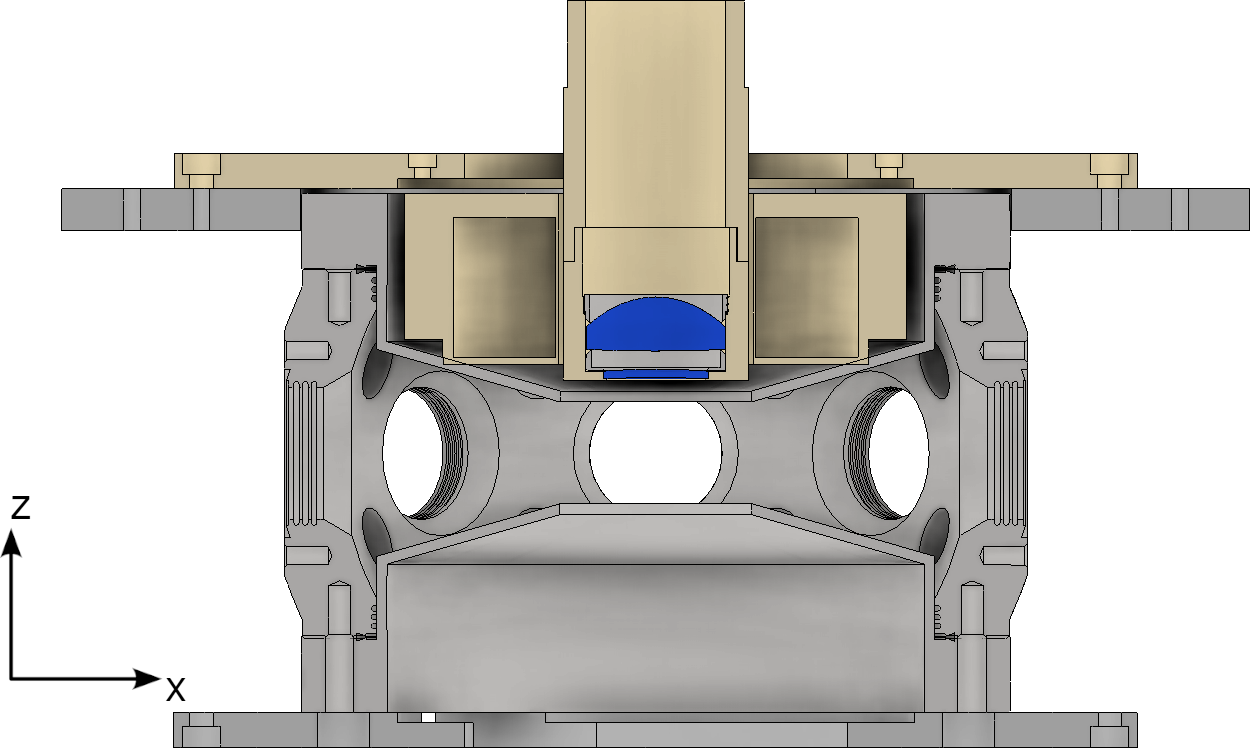}
    \caption{Vertical cut view of the main chamber showing the high-resolution
    microscope placed along the $z$ axis, above the upper vertical viewport. }
    \label{fig:FigSM1}
  \end{center}
\end{figure}

We measure the atomic density distribution of the $^{41}$K-$^{87}$Rb mixture in
the $xy$ plane by means of a high-resolution objective placed, outside the
vacuum system, a few mm above the upper vertical viewport of the main chamber,
which is mounted on a re-entrant CF150 flange (see Fig. \ref{fig:FigSM1}). The
objective, made with a plano-convex aspheric lens and a positive meniscus, was
designed with the help of the commercial ray-tracing software OSLO Educational.
The chosen aspheric lens (Edmunds - 66334) has a diameter of 40 mm and an
effective focal length of 40 mm (at 587 nm). The meniscus lens, custom
manufactured in fused silica, has a diameter of 30 mm and curvature radii of
200 mm and 335 mm, corresponding to a focal length of -1000 mm. To correct for
the aberrations introduced by the viewport, the meniscus is placed in between
the viewport and the lens, approximately at 4 mm from the latter. The optics are
assembled  into a custom-designed tube, made with PEEK polymer,, which is in
turn held by a five-axis optical mount for  alignment adjustment (not shown in
Fig. \ref{fig:FigSM1}). From design, we expected a nominal Strehl ratio
\cite{Mahajan1983} of 0.87 and a radius of the Airy-disk of 1.3 $\mu$m, both at
767 nm and at 780 nm, namely the $^{41}$K and $^{87}$Rb imaging wavelengths. The
performance of the microscope was first characterized on a dedicated optical
table using a replica of the viewport in combination with a 1951 USAF resolution
test chart and a 1-$\mu$m pin-hole. We achieved a resolution of 1.5(1) $\mu$m
at 780 nm. At 767 nm, we observed similar results at a different focal distance,
i.e. approximately 0.02 mm closer to the objective.

In the experiment, the absorption images of both atomic species are recorded by
the same CCD camera: a $^{41}$K image followed by a $^{87}$Rb one. The $^{41}$K
component is imaged \textit{in situ} at the final value of $B_z$. To this end,
we employ the $^{41}$K repumper light, with polarization $\sigma^{-}$ with
respect to the $z$ axis, and a detuning of $-117$ MHz with respect to the
corresponding $F=1 \rightarrow 2'$ transition at zero magnetic field. The
$^{87}$Rb component is instead imaged after a time of flight (TOF) of 2 ms, once
the atoms are approximately in the objective focal plane for the corresponding
wavelength of 780 nm. Rb imaging is performed at zero magnetic field on the
$F=2\rightarrow 3'$ cycling transition after an optical pumping stage.

For the final \textit{in situ} test of the imaging system, we have recorded the
density distribution of a strongly attractive mixture held in an optical dipole
trap. Specifically, we measured the $^{41}$K axial widths and compared them with
the ones expected for the system's ground state calculated under the
experimental conditions. In agreement with the theoretical predictions, we found
that we are able to image high-density $^{41}$K atomic samples with an average
width (RMS) $\Bar{\sigma} = \sqrt{\sigma_x \sigma_y}$ as small as 1.5 $\mu$m.

\begin{figure}[t!]
  \begin{center}
    \includegraphics[width=1\columnwidth]{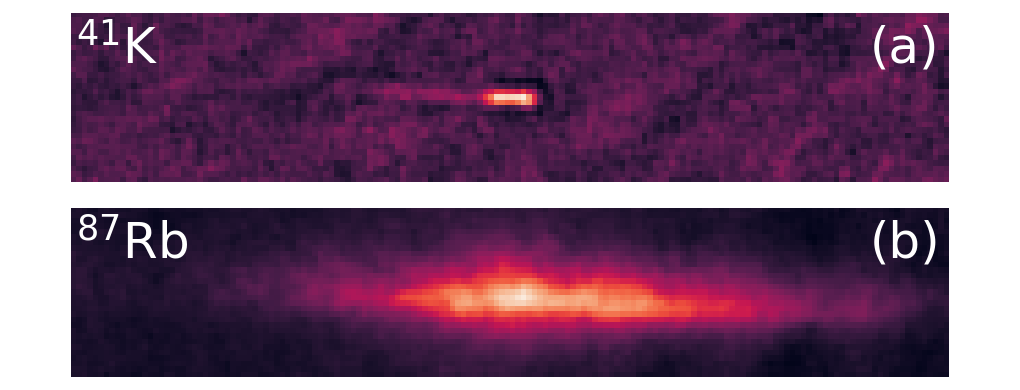}
    \caption{Sequential absorption imaging of $^{41}$K and $^{87}$Rb in the
      $xy$ plane ($238\;\mu$m $\times$ $45\;\mu$m). First $^{41}$K atoms are
      imaged \textit{in situ} at the final value of $B_z$  (a) and then $^{87}$Rb
      atoms are imaged at $B_z=0$ after a TOF of 2 ms (b).}
    \label{fig:FigSM2}
  \end{center}
\end{figure}

As an example, in Fig. \ref{fig:FigSM2} we show absorption images of $^{41}$K
and $^{87}$Rb in the strongly-attractive regime, after an evolution time of 20
ms in the waveguide. We observe that $^{41}$K, the minority component, is almost
bound while $^{87}$Rb, the majority component, is only partially bound. Further,
in the latter case the bound component is not clearly resolved due to the
imaging procedure.

\subsection{\label{sec:level3} Time-of-flight measurements}

\begin{figure}[t!]
  \begin{center}
    \includegraphics[width=0.8\columnwidth]{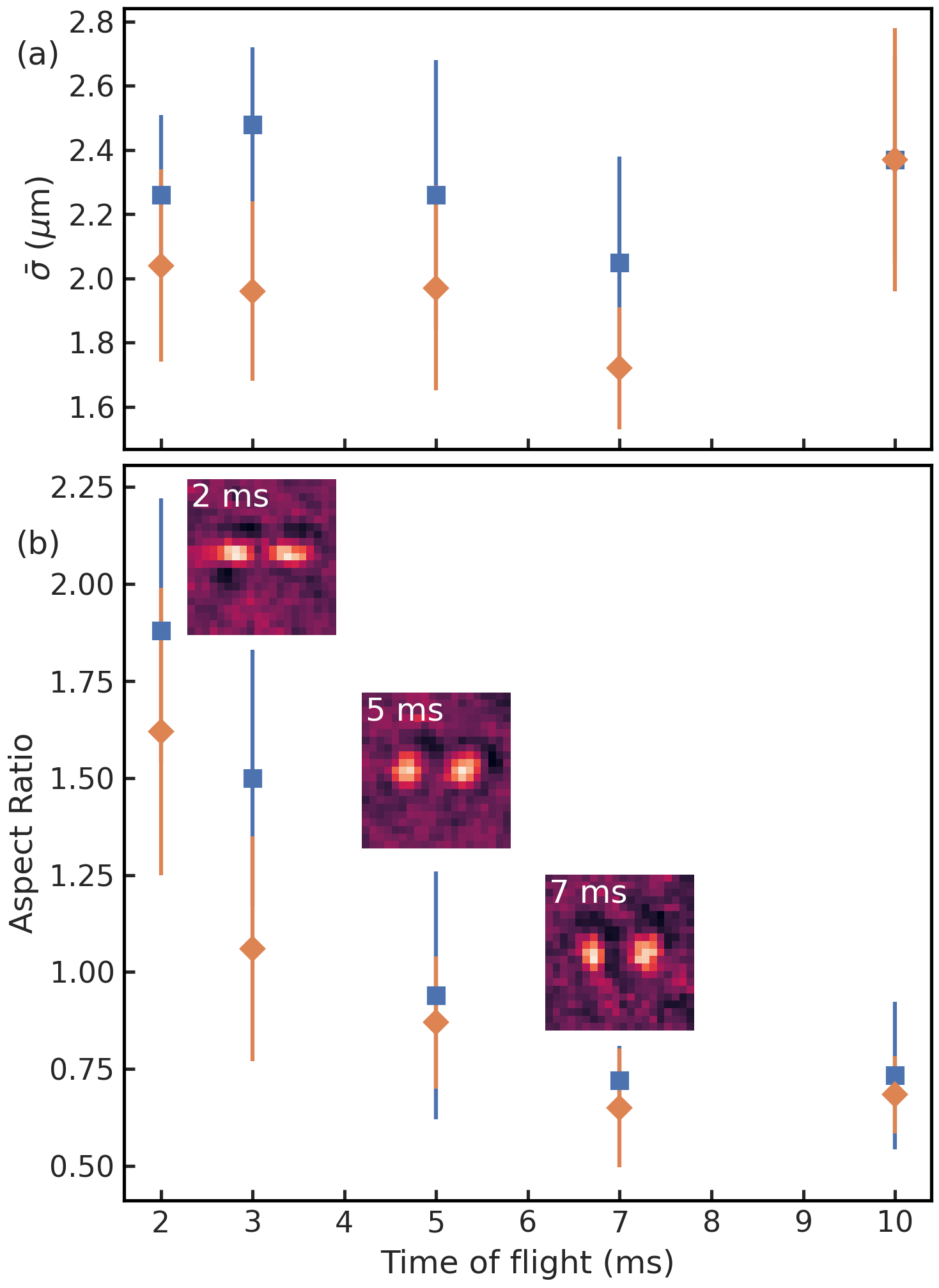}
    \caption{(a) Average size $\Bar{\sigma}$  and (b) aspect ratio of two
      daughter droplets (blue squares and orange diamonds), produced by a breakage
      event in the waveguide, as a function of TOF.  The error bars correspond to a
      standard deviation of typically three independent measurements. The insets show
      typical absorption images of $^{41}$K at 2, 5 and 7 ms of TOF ($30\;\mu$m
      $\times$ $30\;\mu$m). Images are rotated to align them with the $x$ axis.}
    \label{fig:FigSM3}
  \end{center}
\end{figure}

We identify the localized states observed during the dynamics with self-bound
quantum droplets by following their free fall. To this end, we let the sample to
evolve within the waveguide up to 20~ms, when according to \textit{in situ}
absorption imaging it splits into two smaller clusters under the same conditions
of Fig.~1 in the main text. At this point, we suddenly switch off the waveguide
and we record the atomic density distribution by TOF imaging.
Without the optical potential, the atoms fall down under the effect of the
gravity, thus, for each TOF we adjust the focal position of the objective by
means a high-precision motorized vertical translation stage. Fig.
\ref{fig:FigSM3} shows the free-space evolution of the average width
$\Bar{\sigma}$ and the aspect ratio, defined as $\sigma_x/\sigma_y$, of two
fragments produced in the waveguide. We find that both
clusters do not expand, within our imaging resolution, up to a TOF of 10 ms. This
allows us to identify such localized states as quantum droplets, distinguishing
them from solitons, which are stable solutions in the waveguide thanks to the
radial confinement \cite{Cheiney2018}. Furthermore, we observe that the
initially stretched droplets after abruptly switching off the trapping potential
undergo a quadrupole excitation, as shown in Fig. \ref{fig:FigSM3} by the change
of the aspect ratio. 

\section{Numerical Simulations}

We performed numerical simulations by solving two coupled, generalized, time-dependent Gross-Pitaevskii (GP) equations, as detailed below.

The GP energy functional, which includes both the mean-field term and the
Lee-Huang-Yang (LHY) correction accounting for quantum fluctuations in the local
density approximation, is given by \cite{Ancilotto2018}: 
\begin{align}
  E &= \sum_{i=1}^{2}\int \left[\frac{\hbar^2}{2m_i}|\nabla \psi_{i}(\bm{r})|^2 +
  V_{i}(\bm{r},t) n_{i}(\bm{r})\right]d\bm{r} \nonumber + \\
  \frac{1}{2}&\sum_{i,j=1}^{2}g_{ij}\int n_{i}(\bm{r})n_{j}(\bm{r})d\bm{r}
  +\int {\cal E} _{\rm LHY}(n_1(\bm{r}),n_2(\bm{r}))d\bm{r}\,,
  \label{eq:energyfun}
\end{align}
where $m_{i}$ are the atomic masses,  $V_i(\bm{r},t)$ the external potentials, and
$n_i(\bm{r})=|\psi _i(\bm{r})|^2$ the densities of the two components ($i=1,2$).
The LHY correction reads \cite{Petrov2015}
\begin{align}
  {\cal E} _{\rm LHY} &= \frac{8}{15 \pi^2} \left(\frac{m_1}{\hbar^2}\right)^{3/2}
  \!\!\!\!\!\!(g_{11}n_1)^{5/2} 
  f\left(\frac{m_2}{m_1},\frac{g_{12}^2}{g_{11}g_{22}},\frac{g_{22}\,n_2}{g_{11}\,n_1}\right)
  \nonumber \\
  &\equiv \kappa  (g_{11}n_1)^{5/2}f(z,u,x),
\end{align}
with $\kappa=8 m_1^{3/2}/(15 \pi^2 \hbar^3)$ and $f(z,u,x)>0$ being a dimensionless function of the parameters $z\equiv m_2/m_1$, $u\equiv g_{12}^2/(g_{11}g_{22})$, and $x \equiv g_{22}n_2/(g_{11}n_1)$ \cite{Petrov2015,Ancilotto2018}. 
The mixture is characterized in terms of the intraspecies $g_{ii}=4\pi \hbar^2 a_{i}/m_{i}$, 
and interspecies $g_{12}=2\pi \hbar^2 a_{12}/m_{12}$ coupling constants, where $m_{12}=m_{1}m_{2}/(m_{1}+m_{2})$ is the reduced mass. 
The initial stationary configurations are numerically computed by minimizing the above GP energy functional through imaginary-time evolution, using a steepest descent algorithm \cite{press2007} that iteratively propagates the two condensate components. 

The time-dependent GP equations can be obtained through the variational
principle $i\hbar\partial_{t}\psi_i=\delta E/\delta\psi_{i}^{*}$
\cite{dalfovo1999}, yielding
\begin{equation}
  i \hbar {\partial \psi _i \over \partial t} =
  \left[-{\hbar^2 \over 2m_i}\nabla ^2 + V_i + \mu_{i}(n_1,n_2) \right]\psi _i
   \, , 
  \label{eq:gpe}
\end{equation}
with \cite{DErrico2019}  
\begin{equation}
  \mu_{i}\equiv \frac{\delta E}{\delta n_{i}} = g_{ii}n_i+g_{ij}n_j+
  \frac{\partial {\cal E}_{\rm LHY}}{\partial n_i}\quad \,( j \ne i) \,.
  \label{eq:chempot}
\end{equation}
These equations are solved by means of a FFT split-step method (see, e.g., Ref. \cite{jackson1998}). 
During the simulations, the LHY correction is kept always on, regardless of the interspecies interaction. We checked that gradually increasing the LHY term during the ramp in $a_{12}$ does not significantly affect the system dynamics. 
The effect of three-body losses has been included (where indicated) by adding a dissipative term, $-(i/2)\hbar K_3 \int  n_1(\bm{r},t) n_2(\bm{r},t)^2d\bm{r}$, with $K_3=7~\times~10^{-41}$~m$^6/$s, to the energy functional in Eq. \eqref{eq:energyfun}.

We have performed different series of simulations considering for the external potentials $V_i$ either the experimental one (including the two far-off resonance laser beams, the magnetic field used to tune the interspecies interaction, the magnetic gradient used to compensate the gravitational field, and gravity), or its harmonic approximation with concentric traps. 

To simulate the experimental sequence, we first calculate the ground-state of the non-interacting mixture ($a_{12}=0$) in a three dimensional trap. 
The dynamics is then initiated by linearly ramping $a_{12}$ to the final value $a_{12}^f$ within the range $[-90,-82]a_0$. 
Following this, the confinement along the $x$-direction is exponentially ramped down, as in the experiment, allowing the system to evolve in a waveguide. 
The simulations shown in  Fig.~2 of the main text were performed with an unbalanced population of the two components, considering three-body losses that primarily occur during the $a_{12}$ ramp. 
After verifying that the resulting dynamics are very similar when starting with a balanced population and neglecting three-body losses, all other simulations were performed under these latter conditions.

\begin{figure}[t!]
  \begin{center}
    \includegraphics[width=1\columnwidth]{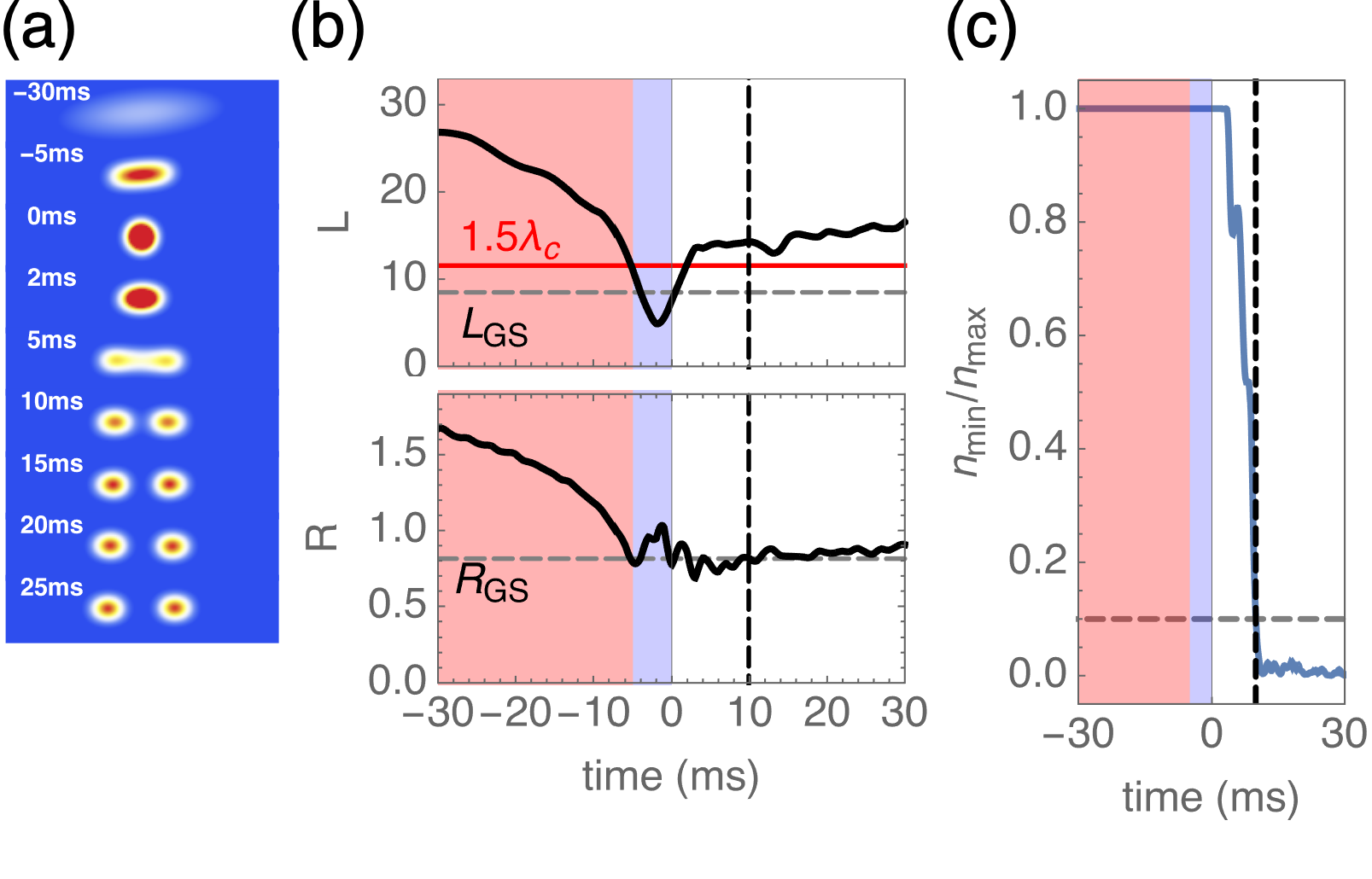}
    \caption{(a) Density profiles integrated along $z$ obtained from the GP simulation with $N_1=N_2=4\times 10^4$, $a_{12}^f=-85\, a_0$, $t_\textrm{ramp}=t_b-t_a=25$ ms, and $t_0-t_b=5$ ms [the times $t_a$, $t_b$, and $t_0$ are indicated in the experimental sequence in Fig. 1(b) of the main text]. 
    The image sizes  are $10\,\mu$m $\times 40\,\mu$m. 
    (b) Solid black lines represent the length $L$ and radius $R$ of the filament as a function of time. The horizontal dashed gray lines show the length $L_{0}$ and the radius $R_{0}$ of the calculated ground state in the waveguide at $a_{12}=a_{12}^f$, while the vertical dashed black line represents $t_{\textrm{break}}$. 
    The solid horizontal red line correspond to $1.5\lambda_c=11.6\,\mu$m. 
    (c) The solid blue line shows the behavior of the ratio $n_{\textrm{min}}/n_{\textrm{max}}$ of the filament as a function of time. 
    The horizontal dashed gray line represents the ratio $n_{\textrm{min}}/n_{\textrm{max}}=0.1$, which we use to determine $t_{\textrm{break}}$, shown as the vertical black dashed line. 
    The red and blue areas correspond to the time intervals during which the interactions change and the crossed-beam confinement is turned off, respectively.}
    \label{SMth1}
  \end{center}
\end{figure}

In Fig. {\ref{SMth1}}(a), we show the density of the $^{41}$K component (with $^{87}$Rb being similar) integrated along the $z$-direction (the experimental field of view), convolved with a Gaussian function accounting for the finite resolution of the experimental imaging system. 
In this simulation, performed for a ramp time $t_\textrm{ramp}=25$ ms, the number of atoms 
in the two components are $N_1=N_2=4\times10^4$, the final interspecies scattering length 
is $a_{12}^f=-85$~$a_0$, and the external potential is the one used in the experiment.

The integrated densities are plotted during the time sequence, starting (for
consistency with the main text) at $t=-30$~ms when $a_{12}=0$. Note that the
horizontal axis corresponds to the direction of the waveguide, and at the
starting time, the cloud is slightly tilted do to the presence of the angled
‘‘crossed beam''. 
At the end of the ramp, at $t=-5$~ms, the two species become strongly attractive, with $a_{12}=-85$~$a_0$. 
At this point, the axial confinement is reduced to zero in 5~ms, while the cloud is still shrinking in the axial direction. 
Then, the filament begins to expand in the waveguide, eventually splitting into two droplets, which exhibit shape excitations.

\begin{figure}[t!]
  \begin{center}
    \includegraphics[width=1\columnwidth]{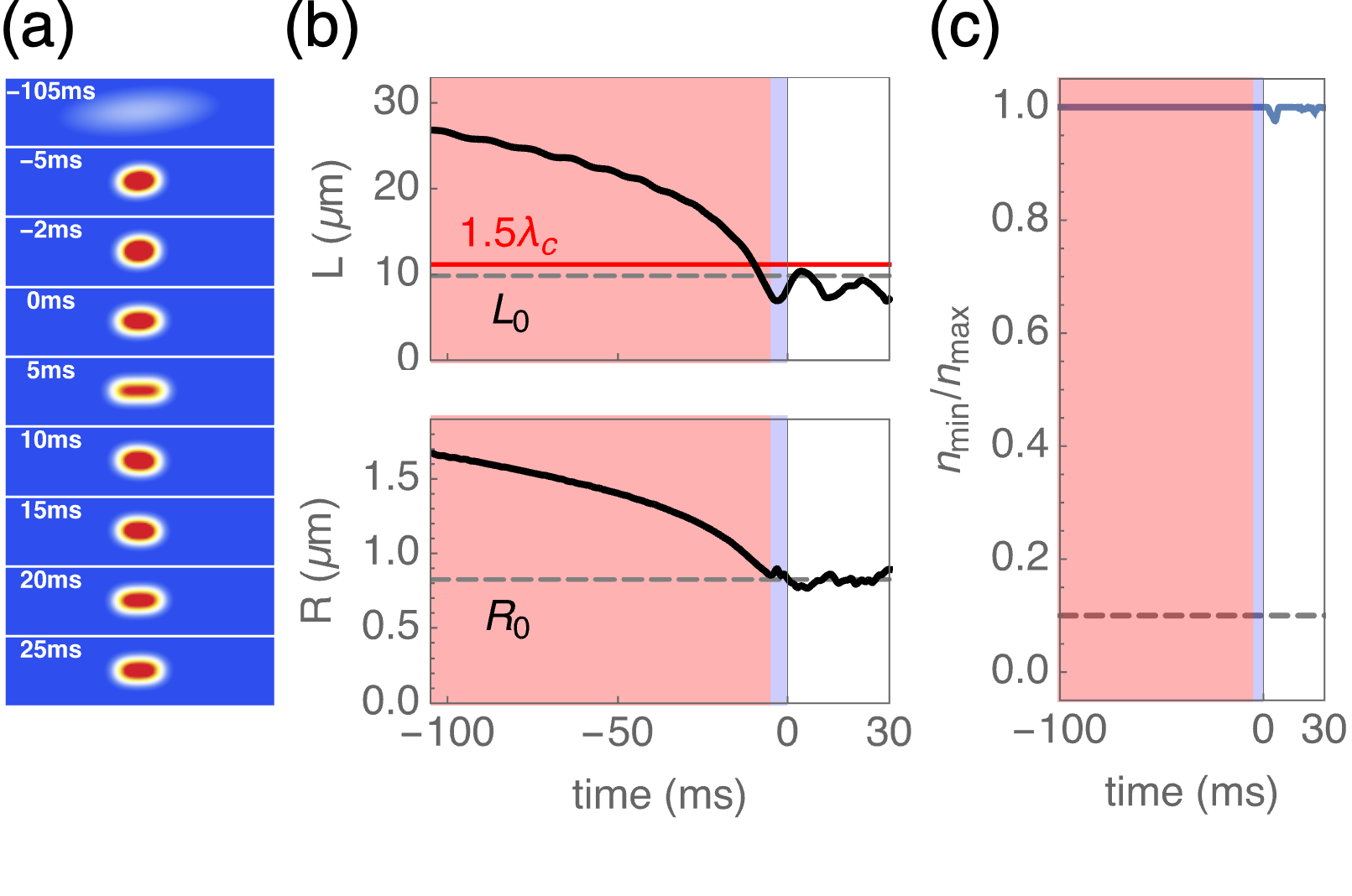}
    \caption{Same as Fig. \ref{SMth1}, but for a longer interaction ramp, $t_\textrm{ramp}=100$ ms. All the other parameters are unchanged. 
    (a) Density profiles obtained from the GP simulation, integrated along $z$.
    (b) Length $L$ and radius $R$ of the filament as a function of time (solid black lines). The horizontal dashed gray lines show the length $L_{0}$ and the radius $R_{0}$ of the calculated ground-state in the waveguide at $a_{12}=a_{12}^f$. 
    The solid horizontal red line correspond to $1.5\lambda_c=11.6\,\mu$m. 
    (c) The solid blue line represents the ratio $n_{\textrm{min}}/n_{\textrm{max}}$ of the filament, while the horizontal dashed gray line indicates the value that  defines $t_{\textrm{break}}$. 
    The red and blue areas correspond to the time intervals during which the interactions change and the crossed-beam confinement is turned off, respectively.}
    \label{SMth2}
  \end{center}
\end{figure}

To better understand the system dynamics, in Fig. {\ref{SMth1}}(b) we show the behavior of the filament length $L$ and radius $R$ as a function of time. 
The length $L$ is defined as the distance along the waveguide between the two endpoints of the density distribution where the density drops to $1/20$ of its maximum value, while the radius $R$ is the mean standard deviation $\sigma$ 
along the radial direction.
The colored areas in the figure represent the time intervals in which interactions are changed (red) and when the confinement of the crossed beam is turned off (blue).
For comparison, the horizontal dashed gray lines represent the corresponding dimensions of the waveguide ground state $L_{0}$ and $R_{0}$ at $a_{12}=a_{12}^f$. 
To evaluate the filament breaking time $t_{\textrm{break}}$, we consider the central density profile of the $^{41}K$ component along  $x$. 
During the evolution, we calculate its maximum $n_{\textrm{max}}$ and its corresponding position $x_{\textrm{max}}$. 
Then, we find the minimum of the density $n_{\textrm{min}}$ in the range $[-x_{\textrm{max}};x_{\textrm{max}}]$.
We consider the ratio $n_{\textrm{min}}/n_{\textrm{max}} (t)$ and define the breaking time as the time when $n_{\textrm{min}}/n_{\textrm{max}}(t_{\textrm{break}})=0.1$; we have verified the results are not very sensitive to this value up to $0.2$. 
Figure {\ref{SMth1}}(c) shows the value of $n_{\textrm{min}}/n_{\textrm{max}}$ as a function of time, represented  by a solid line, while the vertical dashed line [shown both in Fig. {\ref{SMth1}}(b) and (c)] represents $t_{\textrm{break}}$.

As one can see from Fig. {\ref{SMth1}}(b), during the evolution in the
waveguide, the radius performs small oscillation around $R_{0}$, allowing us to define a constant $R$ value (determined by the waveguide confinement) as its mean.
Consequently, assuming the splitting into droplets is caused by capillary
instability, we can calculate the wavelength of the most unstable mode of the
Rayleigh-Plateau spectrum \cite{Eggers_2008} as $\lambda_c= 2 \pi/0.697\,R\simeq 9
\,R\simeq7.7\,\mu$m. In Fig. {\ref{SMth1}b}, we show the value $1.5\lambda_c$
as a horizontal solid red line,  which represent the minimum length to have
a splitting of the filament caused by the density modulation at the wavelength
$\lambda_c$. The number of resulting droplets can be obtained from the filament
length at $t=t_{\textrm{break}}$, which we call $L_{\textrm{max}}$. In the simulation shown in Fig. {\ref{SMth1}}, we have $L_{\textrm{max}}=14\,\mu$m, and a number of droplet $N_d=\lfloor L_{\textrm{max}}/\lambda_c-1/2 \rfloor+1=2$.

In Fig. \ref{SMth2}, we show 
a similar analysis for a longer interaction ramp, $t_\textrm{ramp}=100$ ms.
Remarkably, in this case the filament does not reach the critical value of $1.5 \lambda_c$ and therefore does not break. 
This has also been observed in the experiment, as reported in Fig. 2 of the main text.

\subsection{Systematics}

To systematically study the effect of different values of $N$ and $a_{12}$, 
as presented in Figs.~3 and 4 of the main text, we have performed simulations in a simplified axially symmetric configuration assuming concentric harmonic potentials, to reduce the simulation times.
The initial frequencies were chosen as $\nu_{Rb}=(30,100,100)$~Hz and $\nu_K=1.36 \nu_{Rb}$, providing a reasonable approximation of the experimental potentials.
During the evolution, we simply removed the axial potential. 
We have verified that the dynamics up to the breaking of the filament are substantially equivalent in both the approximated harmonic potential and the potential used in the experiment. 

\begin{figure}[t!]
  \begin{center}
    \includegraphics[width=1\columnwidth]{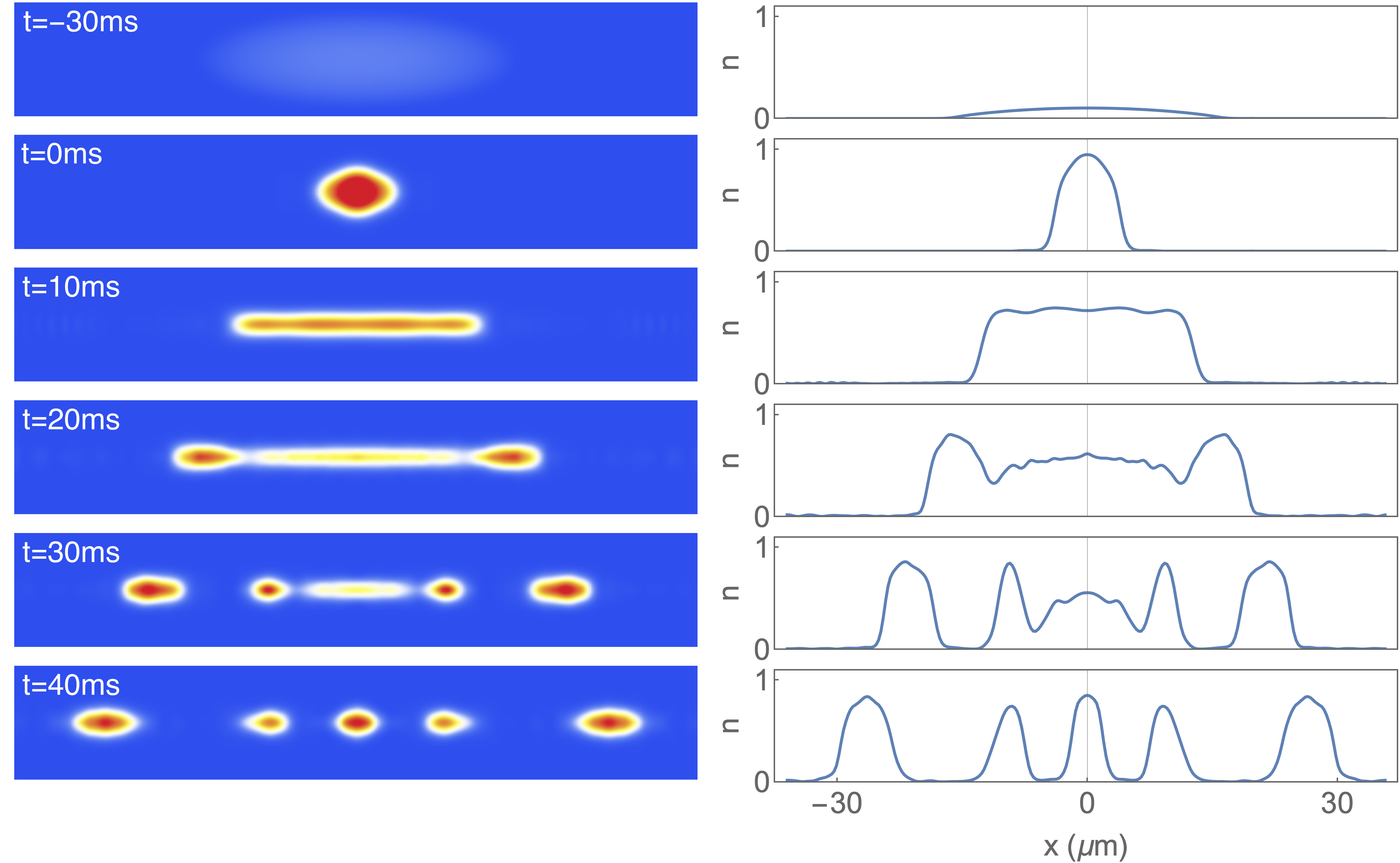}
    \caption{Left: Density profiles obtained from
    the GP simulation with $N_1=N_2=8\times 10^4$, $a_{12}^f=-84\, a_0$ ($t_\textrm{ramp}=t_b-t_a=25$ ms, and $t_0-t_b=5$ ms). Right: 
    Cut of the density profiles along the waveguide direction at $r=0$ (in arbitrary units).}
    \label{SMth3}
  \end{center}
\end{figure}

As an example, in Fig. {\ref{SMth3}}, we show the result of a simulation with $a_{12}=-84\,a_0$ and $N_1=8 \times 10^4$, corresponding to a filament that breaks into five droplets.
The dynamics of fragmentation shows remarkable similarities with the breakup 
of a classical (inviscid) filament, which is usually initiated by the formation of two droplets at the filament end (pinching-off), as discussed by \cite{Castrejon_PRL2012}.
We emphasize that, since the newly formed droplets are highly excited, they undergo large shape oscillations and can come into contact during their evolution, which may lead to variations in their number over time.

\begin{figure}[t!]
  \begin{center}
    \includegraphics[width=.8\columnwidth]{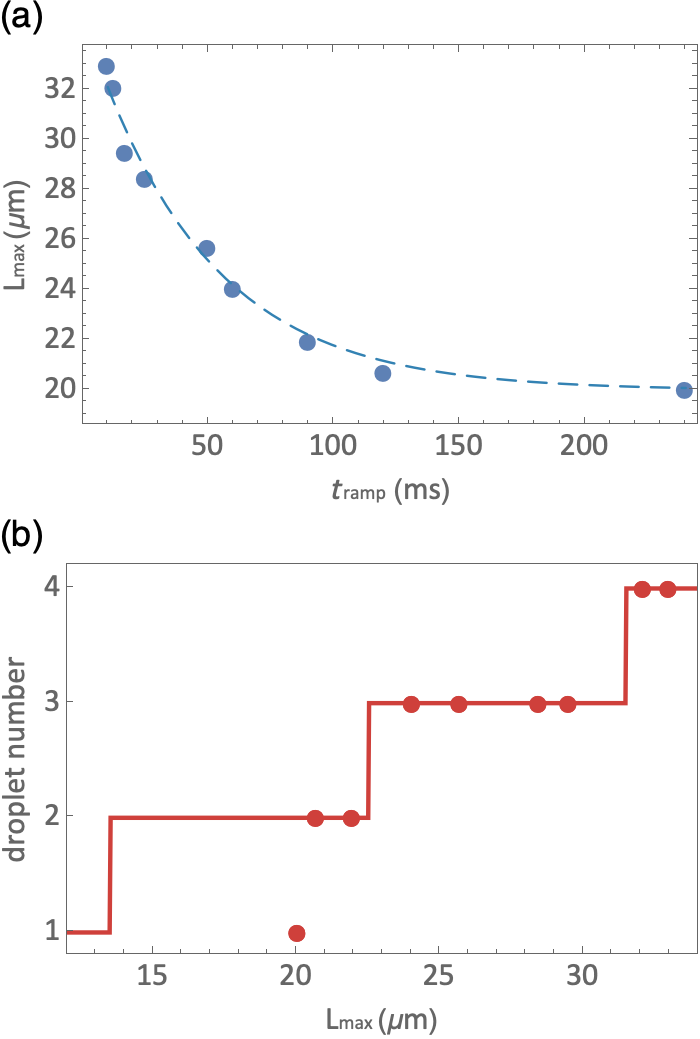}
    \caption{Outcomes of the simulations performed with $a_{12}^f=-84 a_0$ and $N_1=N_2=5 \times 10^4$ for different values of the ramping time $t_{ramp}$: (a) $L_{max}$ versus $t_{ramp}$ (the dashed line is a guide to the eye); (b) number of droplets produced during the filament expansion (data points), and prediction of the PR instability $N_d=1+\lfloor L_\mathrm{max}/\lambda_c-1/2 \rfloor$ (continuous line), here $R=1 \mu$m.  }
    \label{SMth4}
  \end{center}
\end{figure}

To ensure that the presented results are general and do not depend on the specific procedure used to excite the droplet compression-elongation mode, we have simulated the droplet dynamics for a representative case with $N_1=5 \times 10^4$ and $a_{12}^f=-84$~$a_0$, while varying only the duration of the interaction ramp $t_\textrm{ramp}=t_b-t_a$. We find that, if $a_{12}$ varies rapidly enough, the droplet axial size exceeds the critical length and the droplet possibly breaks up. 
The results are shown in Fig. \ref{SMth4}. Panel (a) shows that $L_{max}$ increases as the duration of the interaction ramp is decreased. In panel (b) we compare the number of droplets from the simulations with those predicted by the relation $N_d=1+\lfloor L_\mathrm{max}/\lambda_c-1/2 \rfloor$.
From this analysis, we conclude that our interpretation in terms of capillary instability does not depend on the specific timing of the experiment. Indeed, the model reliably predicts the outcome of the breakup process, while the emergence of small discrepancies for long $t_{ramp}$ (short $L_{max}$) can be attributed to the dynamical nature of the process and the finite length of the filament.

We also performed a detailed investigation of ground-state properties of the mixture in the waveguide. 
This included examining its length and radius as a function of $a_{12}$ and $N$, as well as analyzing the droplet character of the ground state. 
The latter was accomplished by simulating its time-of-flight free expansion and identifying the droplet region as the parameters region where the radial size of the ground state in free space does not expand but instead oscillates.  
This analysis has been used to support the overall theoretical framework presented in this work.

\bibliography{biblio}